**Terahertz Devices Using the Optical Activation of GeTe Phase Change Materials: Towards Fully Reconfigurable Functionalities**

*Maxime Pinaud, Georges Humbert, Sebastian Engelbrecht, Lionel Merlat, Bernd Fischer, Aurelian Crunteanu**

M. Pinaud, Dr. G. Humbert, Dr. A. Crunteanu

XLIM Research Institute,

CNRS/ University of Limoges

Limoges 87060, France

E-mail: aurelian.crunteanu@xlim.fr

Dr. S.Engelbrecht, Dr. L. Merlat, Prof. B. Fischer

ISL, French-German Research Institute of Saint-Louis,

Saint Louis, 68300, France

**Keywords**

Phase-change materials, optical activation, reconfigurable terahertz metamaterials, polarizer

**Abstract**

Reconfigurable multifunctional terahertz (THz) devices are crucial for the development of practical applications such as high-speed communications, spectroscopy and imaging, etc. but their expansion is still requiring efficient agility functions operating in the THz domain. Chalcogenide phase change materials (PCMs) with their broadband response, non-volatile and fast reversible transitions between a dielectric and metal-like phases, were successfully investigated as agile elements in photonics or electronics applications but their potential for controlling the propagation of THz waves is still under evaluation. Here, we are demonstrating the optical control of a specific state of the germanium telluride (GeTe) phase change material and its integration as control element for enable future efficient optically reconfigurable THz devices. The excellent contrast of the material's THz electrical properties in the two dissimilar states were used for optical-induced fast modulation of THz resonances of a hybrid metamaterial based of arrays of split ring resonator metallic structures integrating GeTe patterns. We experimentally confirm for the first time the feasibility to develop all dielectric (metal free) GeTe-based THz polarizers presenting a broadband response, a high extinction ratio when the GeTe is in the metal-like phase (up to 16.5 dB) and almost transparent when the material is in the amorphous phase. The presented highly functional approach based on non-volatile, optically controlled multi-operational THz devices integrating PCMs, is extremely stimulating for generating disruptive developments like field- programmable metasurfaces or all-dielectric coding metamaterials with multifunctional capabilities for THz waves manipulation.

## 1. Introduction

Terahertz (THz) waves hold very promising applications such as high-speed wireless communication, security screening, chemical identification and non-destructive sensing.[1] A prerequisite to the extensive development of these domains at THz frequencies is the development of reconfigurable devices (filters, spatial light modulators, lenses, frequency selective surfaces,…), which allow controlling the properties of electromagnetic waves such as their amplitude, phase, polarization and/ or spatiotemporal distribution of their wave fronts[2]. Reconfigurable devices are essential in reducing the cost, complexity and overall dimensions of THz systems. Reconfiguration strategies and the associated technologies already employed in the optical and microwaves domains are difficult or ineffective to transpose to THz frequencies[3]. Current approaches for manipulating THz radiation are based on artificially engineered metasurfaces or metamaterials consisting of two-dimensional arrays of sub-wavelength metallic resonators[2]. The topology or the electromagnetic properties of these individual building blocks can be modified by the integration in their design of different agility functions such as semiconductors,[4, 5] liquid crystals,[6-8] MEMS devices,[9] phase transition materials ($VO_2$)[10, 11] or graphene [12, 13] which has been proven as effective means to control the properties of THz waves.

However, most of these agile elements have volatile-type responses requiring permanent optical or electrical bias, and complex integration technology with electrical/ optical control schemes over large areas. At the same time, they have to be integrated within the fixed topology of metadevices's for active manipulation of THz waves.

Chalcogenide phase change materials (PCMs) such as GeTe (germanium telluride) or $Ge_2Sb_2Te_5$- GST (germanium antimony telluride) have been successfully integrated for realizing reconfigurable electrical or optical devices. The reconfiguration capability is based on the unique properties of PCMs undergoing a non-volatile phase transition on nanoseconds timescales from an amorphous (insulating) state to a crystalline (metallic), drastically changing their electrical and optical properties upon application of thermal, electrical or optical stimuli. [14-17]

These PCM capabilities have been exploited over the years for realizing non-volatile memories by stacking the material between two electrodes and changing its state using pulsed current-Joule heating [18, 19]. More recently, GeTe and GST compositions have shown their ability to be used as high performance, bi-stable switching devices in the microwaves/ millimeter waves domains, holding high-promises for future multi-reconfigurable telecommunication systems[20, 21]. Their implementation requires a planar electrodes indirect heating scheme using short

voltage/ current pulses applied on a thin-film heater isolated through a dielectric film underneath the PCM films [20-22].

Compact optical metadevices based on PCMs, such as flat optics-like switchable/ tunable filters, multi-focus Fresnel zone plates, super-oscillatory lenses with subwavelength focus, and optical holograms have also been demonstrated [23-28]. Furthermore, optical activation of PCMs between the two states has been widely used for rewritable optical disks storage memories[29, 30]. Nevertheless, the large area capabilities of this technology are based on the association of mechanical movement of the storage medium and activation regions of PCMs limited by the laser beam size.

Most recently, it has been proven that PCMs can be used to realize all-dielectric, metal-free agile functions in the optical domain using optical control of GST-based metadevices.[25, 31] Unlike electrical-based control schemes[18-21], fast optical activation using femtosecond and nanosecond lasers of PCMs allows a more accurate control of the specific thermal diffusion effects associated with the crystallization and amorphization processes during the phase change mechanisms of the materials.[32-34]

In the THz domain, the potential of PCMs as agile elements for controlling the properties of THz waves is still under evaluation. The few published studies report on their THz properties [35], and on the possibility to integrate GeTe and GST stripes within THz metadevices which can adapt their response using thermal or electrical stimulus to change PCM's state only from amorphous to crystalline.[36]

Recently, a THz metadevice based on 2x2 panels of metallic asymmetric split ring resonators (A-SRR) with dissimilar dimensions fabricated on a GST film has been reported to present multilevel states at THz frequencies using the electrical control (through a direct Joule heating scheme) of the crystallinity degree (and thus of the conductivity) of the GST material.[37] While the electrically induced nonvolatile and multi-state modulation of the THz transmission of the devices at multiple frequencies was validated for the amorphous-to-crystalline state of the GST material, the reverse operation of the system corresponding to the crystalline-to-amorphous state change of GST is yet to be proven. Electrical activation has obvious and numerous advantages for individual devices. However, the amorphization/ crystallization processes requires extremely localized areas of PCMs inserted between the command electrodes making it hardly compatible with design constraints of metadevices at THz frequencies.

In addition, the use of subwavelength metallic resonators as elementary building blocks of THz metadevices strongly limits the possibility to achieve large reconfiguration functionalities since the integration in their topology of active reconfigurable elements over wide areas (and associated electrical control) remain complex.

Although PCM-only metadevices made on GeTe-based split ring resonator arrays have been proposed in the past, they show very weak changes in their specific response at THz frequencies during GeTe phase transition from amorphous to the crystalline phase. [36]

Nevertheless, THz devices fabricated with PCM active layers that are compatible with an optical activation between their dissimilar states, would lead to improved performances, simplicity of integration and may attain multiple reconfiguration functions.

Here, we demonstrate for the first time to our knowledge an implementation scheme of PCMs and their associated optical control, foreshadowing the development of PCM-only THz devices with nonvolatile and reversible reconfiguration capabilities.

We validated the fabrication of GeTe phase change films with high contrast of their THz properties between amorphous and crystalline states allowing the first developments of all-PCM THz components. Furthermore, we experimentally show that these PCM films can be successively optically switched on large areas between their amorphous and crystalline phases using nanosecond laser pulses, while keeping an excellent contrast of their THz properties in the two dissimilar states.

These performances were further used for demonstrating optical-induced THz response of a hybrid THz metadevice based of arrays of split ring resonator metallic structures integrating

GeTe patterns. The significant amplitude contrast of the THz transmission between the two states of the GeTe film was finally exploited for developing all-PCM THz components by experimentally demonstrating optically activated GeTe-only THz polarizer devices with compelling performances to incident THz waves. This proof of principle based on the GeTe building elements and their associated activation method holds high-promises for future THz devices with fully reconfigurable transfer functions.

## 2. Thermal and optical activation of GeTe phase change material and its electrical performances at THz frequencies

We investigated the modification of electrical properties of GeTe films obtained on c-cut sapphire substrates, at DC and THz frequencies, for both direct plate heating or optical activation of phase changes (**Figure 1a**) (see also experimental section for details). As fabricated, the 500-nm thick GeTe film is amorphous with high DC resistivity of 800 Ω·cm. In the case of direct heating, the film's two-point resistance is gradually reducing with the temperature until ~175°C, when it is abruptly decreasing and stabilizes to the crystalline state resistance, as illustrated in **Figure1b**. The material retains its conductive state crystalline state[38] with a resistivity value of 12x $10^{-5}$·Ω·cm, upon further annealing up to 300°C and subsequent cooling at ambient temperature, confirming its non-volatile electrical transition.

The GeTe phase change through thermal heating is inducing a large modification of the material's THz properties. The transmitted temporal signals for both amorphous and crystalline 500-nm thick GeTe film recorded by terahertz time domain spectroscopy (THz-TDS) measurements are represented on **Figure 1c**. The corresponding Fast Fourier Transform (FFT)-calculated frequency domain of THz transmission of the GeTe film in amorphous and crystalline states (spectra normalized by the substrate THz transmission) show normalized transmissions around 90% for the amorphous phase and of 1.7% for the crystalline state, with relatively constant values over the entire measured domain (**Figure 1d**). The metallic behavior of the crystalline phase of GeTe is manifested by a strong reflection of the incident THz radiation and therefore a drastically decrease of the amplitude of the transmitted THz pulse.

The thicknesses of investigated GeTe films are much smaller compared to the THz wavelengths and, taking also into account that the GeTe in the amorphous phase is almost transparent to the THz waves, it is very difficult to evaluate the complex permittivity of the 500nm-thick GeTe films prepared in the two states, within acceptable errors.[39]

Still, the analysis of similar THz-TDS experimental data obtained for 1-µm thick GeTe films on sapphire substrates (supplementary material, **Figure S1**) allowed us to extract conductivity values of 1000 ±20% S/m for GeTe in the amorphous phase and (2.2 ±10%) x$10^5$ S/m for GeTe in the crystalline phase in the 0.2-1 THz frequency domain. These values are consistent with previous reported data on different phase change material compositions [35].

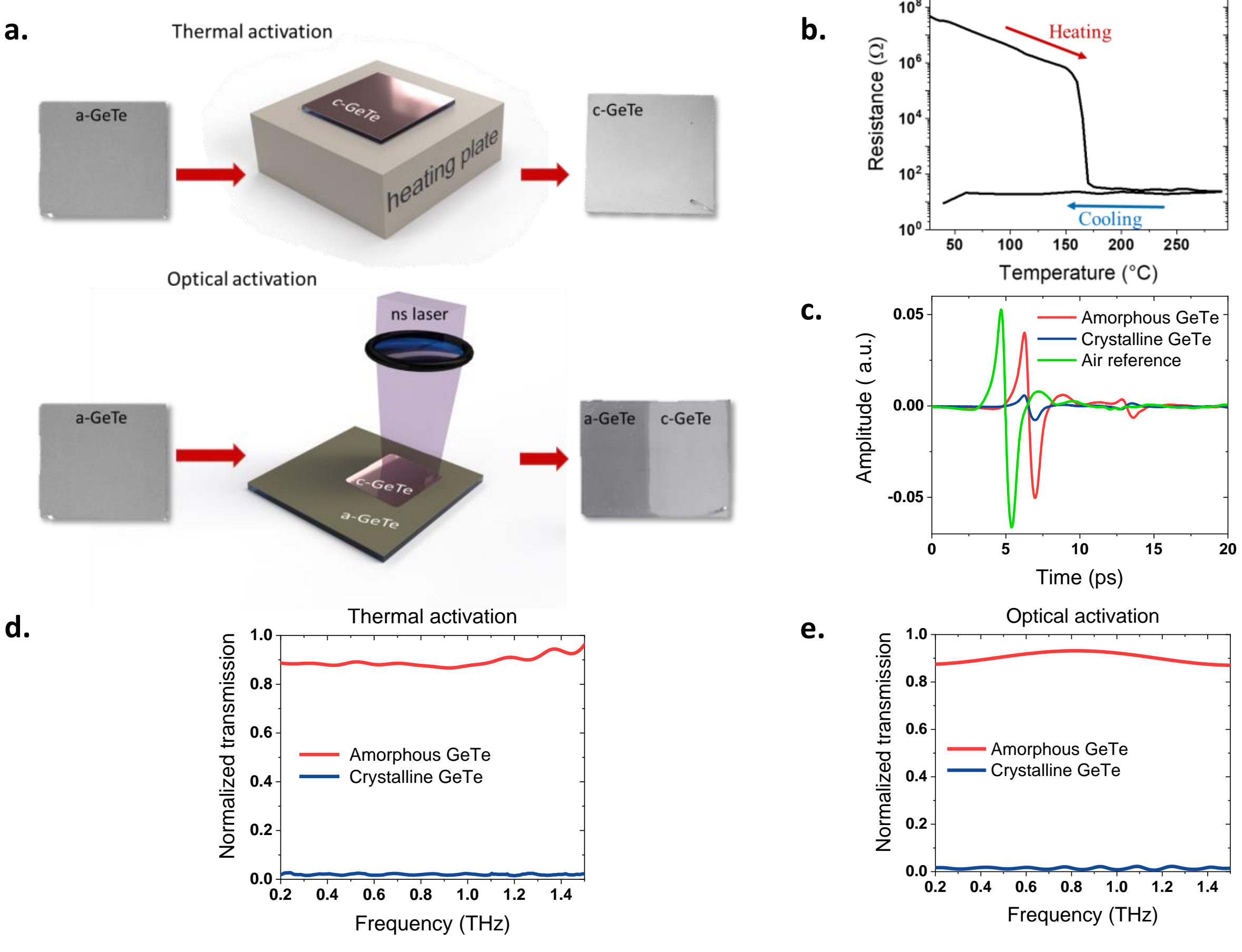

**Figure 1.** (a) Amorphous-to-crystalline phase change of a 500-nm thick GeTe (initially amorphous) film obtained on a sapphire substrate (20x 20 $mm^2$ overall dimensions) using either direct-heating at 300°C on a heating plate or, alternatively, by optical irradiation with nanoseconds laser pulses. (b) Typical resistance variation with temperature (two-probe measurements) of an initially amorphous GeTe film recorded during direct heating experiments. (c) Temporal transmission of THz-TDS signals obtained for the amorphous and respectively, crystalline GeTe films and (d) associated normalized THz transmission of the amorphous and crystalline GeTe phases obtained by direct heating plate thermal activation and (e) equivalent results of amorphous and crystalline GeTe phase obtained using optical activation.

An alternative and suitable method to achieve the amorphous –to-crystalline phase change in PCMs is their optical irradiation using nanosecond laser pulses (**Figure 1a**)[25, 29, 30, 32]. Compared to the plate heating method for which is extremely difficult to cycle in a reversible way the material between its dissimilar phases, the laser-induced phase change in PCMs has proven to be a method with fast switching capabilities, allowing the repetitive and sequential preparation of the material in a specific, desired state [40].

Our approach for optically switching the GeTe layers between the two states is employing the direct irradiation of the films using 35-ns long UV pulses from an excimer KrF laser operating at 248 nm (**Figure 1a**) [17]. The normalized THz transmission of the optically-crystallized GeTe shows performances identical to those of the crystalline GeTe films prepared by the plate heating method (**Figure 1e**). The employment of the KrF excimer source for laser-induced phase change in GeTe has a major advantage: the convenient way of preparing a specific state of the material in a recurrent and reproducible manner over large areas (the imprint - beam dimensions on the irradiated surface- of the laser beam covering a surface as large as 1.5 x0.5 $mm^2$) by simple direct irradiation or using raster scanning.

Thus, to crystallize an initially amorphous GeTe 500-nm thick layer, an UV laser pulse with a power density of 1.1 x $10^{17}$ W/m$^3$ (peak power volume density) is required to heat the material above its crystallization temperature, $T_C$ (**Figure 2a**). For bringing back the crystalline GeTe film to its amorphous phase, a higher power laser pulse should be employed, with power densities around 8.4 x $10^{17}$ W/m$^3$, in order to rapidly heat the material above its melting temperature $T_M$ and quench it in a disordered (amorphous) phase (**Figure 2b**). This melt-quenching step is essential for reversible change of PCMs between their two states and requires intense and fast thermal pulses (electrical or optical induced) which can be difficult to implement using usual thermal heating approaches involving heating plates or ovens[40]. The finite elements multiphysics simulations indicate that the temperatures rise within the GeTe layers subsequent to the application of the two energy-dissimilar laser pulses (black curves in **Figures 2 a and b** and temperature distribution over the entire laser pulse area on supplementary **Figure S2**) makes possible the reversible change of the GeTe film between the amorphous and crystalline phases. As the optical control of a specific phase of the GeTe material on a large area is critical for the realization of reconfigurable THz devices, we experimentally validated for the first time the reversible amorphous-to-crystalline phase changes of a GeTe layer obtained initially in the amorphous state on a sapphire substrate by recording its two-probe sheet resistance change subsequent to the application of alternating UV laser pulses with distinctive energies (**Figure 2c**).

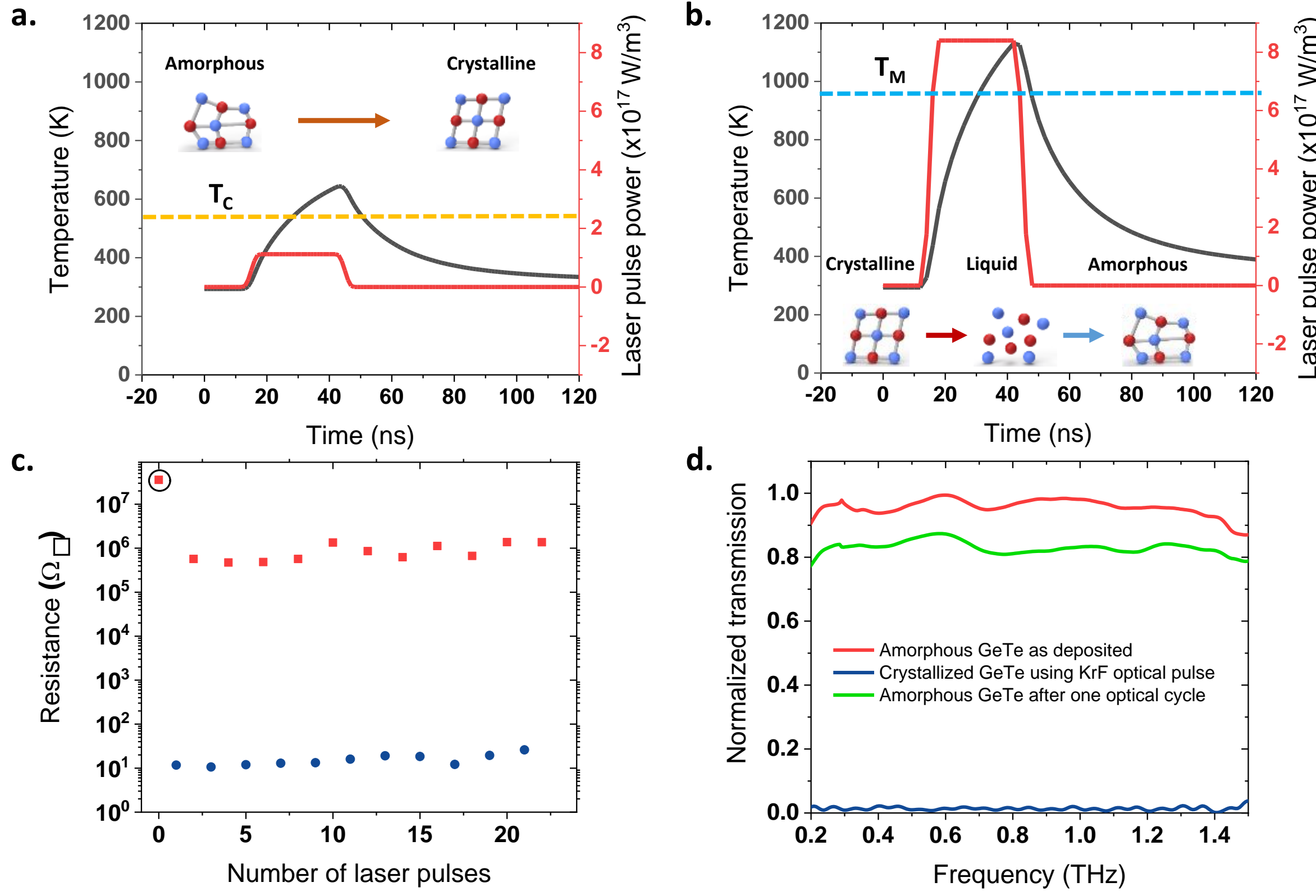

**Figure 2**. a and b. Finite element simulations of temperature rise within a 500-nm thick GeTe film obtained on a sapphire substrate (black curves) during optical phase change induced by nanosecond laser pulses with dissimilar pulse powers (red curves) for **(a)** the crystallization process where the temperature in the initially amorphous film exceed the crystallization temperature ($T_C$) and for **(b)** the melt-quenching process for which the higher-power laser pulse is rising the GeTe temperature above its melting temperature ($T_M$) followed by rapid cooling to a disordered, amorphous phase. **(c)** Four-probes **s**urface resistance (Ω□ or square resistance) variation measurement of the irradiated film submitted to optical irradiation with 20 laser pulses with alternating power energies corresponding to the processes described in Figures 2 a and b and **(d)** THz-TDS measured normalized transmission variation for one optical activation cycle (amorphous- crystalline- melt-quenched optical transformation).

The evolution of the sheet resistance of the film following the sequential transformation from amorphous to crystalline states shows that crystallization and melt-quenched processes are reproducible and that the resistance values in each of the states are stable after more than 20 laser pulses. The initial, as-obtained, amorphous sheet resistance value of the film is higher than the values of the melt-quenched process which can be explained by the laser irradiation process performed in ambient atmosphere, without any capping or protective layer which would have minimized the interaction of the irradiated film with the atmosphere during the melting process and prevent the partial oxidation of the near surface of the film which affect its electrical performances. Nevertheless, the electrical contrast between the resistances values of the crystallized and melt-quenched phases is still higher than $10^5$, resulting in a high THz transmission ratio between the two phases (**Figure 2d**). Thus, at 1 THz, the mean normalized transmission of the optically crystallized GeTe layer is 1.5% and of 83% for the optically melt-quenched, amorphous phase.

In the following sections we will experimentally demonstrate that this optically-prepared high transmission ratio between the two dissimilar states of the GeTe film may be employed not only for recurrent non-volatile switching of THz waves but also for the design of essential terahertz devices with high compatibility to reconfigurable functionalities for which the integration of active GeTe material is straightforward.

## 3. Integration of GeTe in hybrid reconfigurable metamaterial THz device

We exploit the optical activation and the high contrast of THz transmission between amorphous and crystalline phases of GeTe to demonstrate optical activation of a hybrid metamaterial acting as a notch filter in the THz frequency domain. The design of the device is based on a periodic collection of metallic square-shape split-ring resonators (SRR) for which the gap is closed by a GeTe pattern (**Figure 3 a, b**). Such metamaterial SRR-based topologies had proven their capacity to effectively manipulate the THz waves [41, 42], having sharp LC-type resonances for which the frequency and resonance amplitude can be conveniently adjusted by varying their dimensions, shape or and the dielectric environment in the gap of the structure[43, 44].

The simulated response of the proposed device based on a network of SRR unit cells (**Figure 3b**) spaced by 20 µm in both directions in the plane of the overall structure is shown on **Figure 3c**. When the GeTe pattern within the SRR gap is amorphous, it is almost transparent to the THz waves and the normal LC resonance of the SRR structure occurs at around 0.26 THz accompanied by higher order resonances at 0.73 and around 1.1 THz (red curve on **Figure 3c**). When the GeTe is crystalline, it electrically closes the gap of the SRRs and the LC resonance vanished, and different resonance appears at 0.58 THz, corresponding to the mutual coupling between each of the thus modified SRR structures. The SRR metamaterial based on the hybrid integration of GeTe-metallic structures was further fabricated on a c-cut sapphire substrate (**Figure 3d**) and its performances were evaluated using the THz-TDS measurement technique in both configurations (GeTe patterns in the amorphous states and in the optically-crystallized one, respectively). For crystallizing the GeTe integrated in the SRR unit cells, the entire metamaterial structure was irradiated using a single laser pulse, as mentioned before.

The measured normalized THz transmissions of the fabricated metamaterial device are represented in **Figure 3e** for both amorphous and crystalline GeTe in the gap of the SRR structures and they are corresponding very well to the theoretical performances.

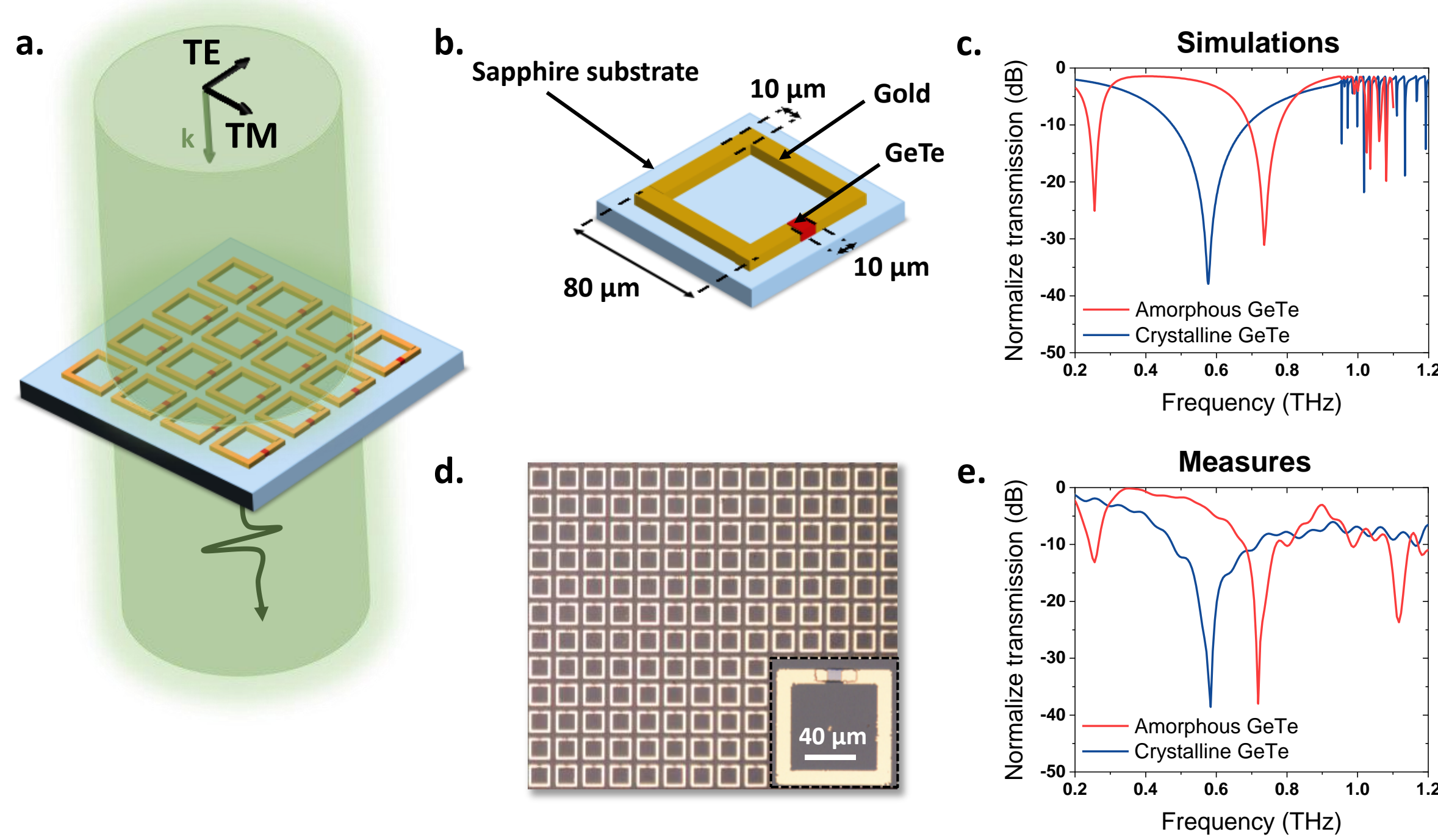

**Figure 3. (a)** Schematic of the THz set-up for evaluating the performances of a metallic split-ring resonators array integrating GeTe patterns prepared under both amorphous and crystalline phases and **(b)** the unit cell parameters of the designed SRR metadevice. **(c)** Simulated THz transmission of the SRRs structure for both configurations of the GeTe within the unit-cell gap (amorphous and crystalline). **(d)** Optical microscopy image of the fabricated SRR device with a close-up look at one unit-cell and **(e)** the measured normalized THz transmission of the SRR device with GeTe patterns in the amorphous and crystalline phases.

Although some differences between the simulated and experimental results are still occurring, they can be explained by the imprecision in evaluating the exact electrical and optical properties of such very thin GeTe films and the slight dimension differences occurring in the microfabrication process between the simulated structure and the fabricated device. The performances of the fabricated device can be also evaluated in terms of the quality factor of each of the resonance peaks, defined as $Q = f_0/\Delta f$, where $f_0$ is the resonance frequency and $\Delta f$ is its bandwidth at half minimum. When the GeTe is amorphous, Q= 9.36 (LC resonance at 0.26 THz), while Q = 15.8 when the GeTe is transformed to its crystalline state (resonance occurring at 0.58 THz). These quality factors are similar to the ones corresponding to metallic-only SRR planar devices (replacing with air the amorphous GeTe and with pure gold patterns the crystalline GeTe in the corresponding designs). Although these figures are also design-dependent, the introduction of GeTe material within the SRR unit structure does not alter their values, implying that the losses induced by the integration of GeTe in the amorphous state or the lower conductivity of crystalline GeTe (worst case scenario when compared to a pure metal conductivity) may be alleviated in such metamaterial hybrid implementations.

The same analysis was performed on identical devices fabricated on silica ($SiO_2$) substrates (see supplementary **Figure S3**) and shows comparable performances, although the specific resonances of the device are appearing at different frequencies, given the different substrate permittivity.

We obtained quasi-identical responses for similar devices integrating GeTe patterns which were prepared to the crystalline state using direct thermal activation on a heating plate (as indicated on supplementary **Figure S4**) which is clearly indicating that the optical-induced phase change of GeTe within the metamaterial structure is a highly effective and much more appropriate method in modifying, on-demand, the THz response of such devices.

The strong variation of the transmission at 0.58 THz (higher than 35dB) of the GeTe implemented SRR device between the optically-induced two dissimilar phases is a strong representation of the capabilities of these type of materials and associated switching technique in the development of low consumption, highly integrated THz devices relevant for switching functions, filters etc. with a plethora of applications in telecommunication and imaging[45].

## 4. All-dielectric reconfigurable THz polarizer based on GeTe phase change material

The strong contrast in THz amplitude signals transmitted through the GeTe films optically prepared in the two different states suggests that simple structures made only in GeTe (without additional metallic structures) could be feasible, offering interesting and promising opportunities for developing active THz reconfigurable devices. In particular, THz polarizers are essential devices, with critical roles in imaging and wireless communication applications [46-49]. For demonstrating the potential capabilities of the GeTe material for these applications, we develop what we believe being the first all-PCM THz polarizer, i.e. the first device using the metallic/ crystalline state of GeTe to produce the selection of the polarization state of a transmitted THz wave.

The design of the proposed polarization device is based on GeTe wire-grid parallel structures on a sapphire substrate and have reconfiguration capabilities (turning on or off its polarization behavior) brought by the optically-active GeTe material (Figure 4).

We designed three different polarizers with dissimilar wire widths (w) and periods (P) which were fabricated by patterning a 500-nm thick GeTe layer on a sapphire c-cut substrate (**Figure 4b**). The polarizer P1 was made with a period P = 10 µm and a wire width w = 3 µm (P1(10/3 on **Figure 4b**). The P2 device was realized with a smaller period P = 6 µm and the same wire width w = 3 µm while for the third design, the width of the wires was reduced to w = 2 µm and the period was fixed to P = 9 µm (denoted as P2(6/3) and P3 (9/2), respectively, on **Figure 4b**). A fourth polarizer structure was realized using gold instead of GeTe to serve as a reference, with P = 6 µm and w = 3 µm(P4(6/3) on **Figure 4c**).

When the GeTe is amorphous, the wire-grid structures act as a transparent dielectric material, allowing TM and TE THz waves to be transmitted with identical results (see supplementary **Figure S5**). However, when the GeTe is brought to its crystalline, metal-like state using UV laser pulses, it induces a wire-grid polarizer effect, blocking a significant part of the TE-polarized waves[43]. This behavior is clearly visible on the simulated and measured performances of the GeTe-based devices shown on **Figure 4c**.

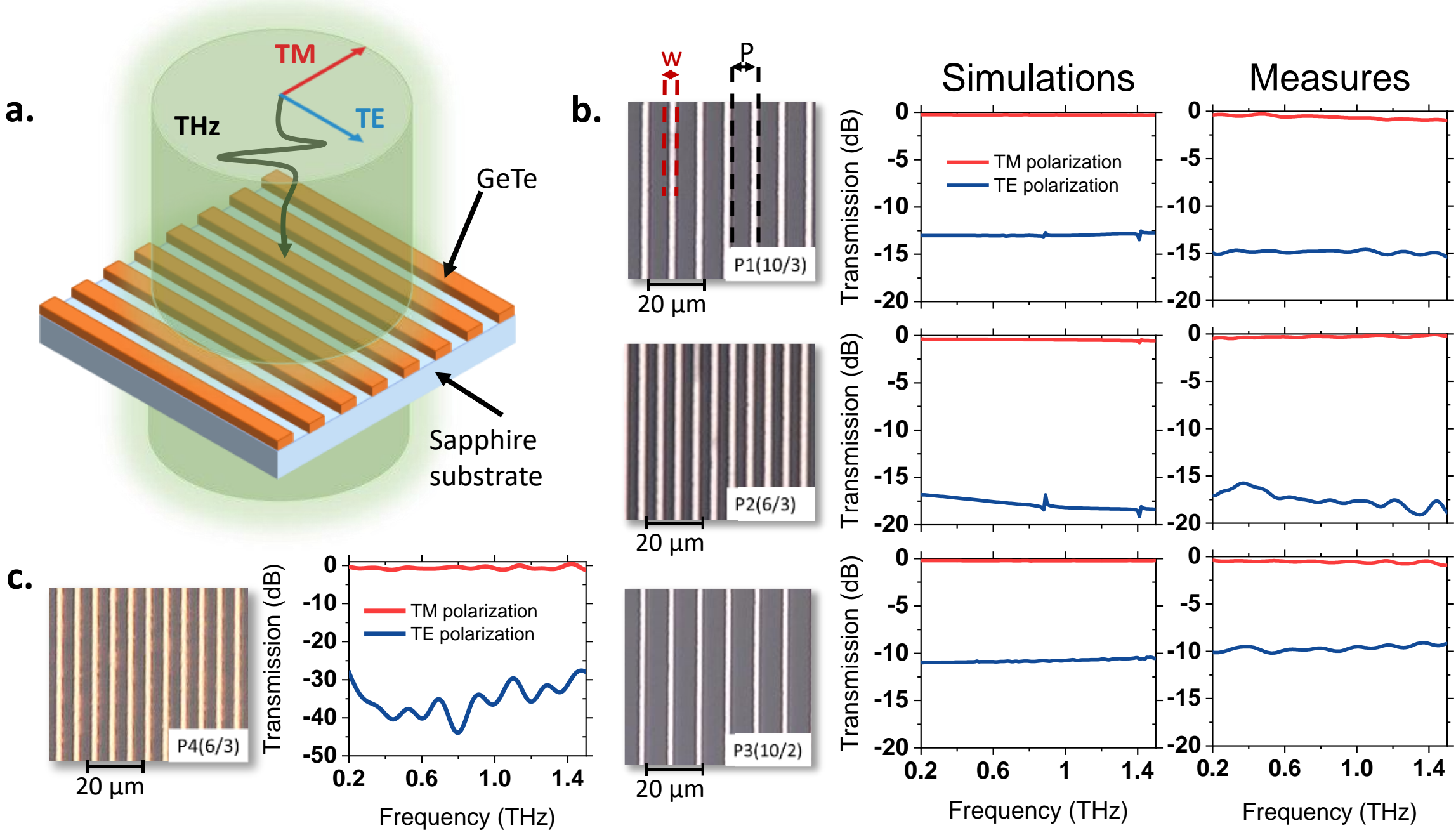

**Figure 4.** **(a)** Measurement scheme of the GeTe-only THz polarizer **(b)** Close-up optical images of the fabricated polarizers with GeTe in the crystalline state (P1, P2, P3) having different line periods, P and line width w (P /w in µm) and their corresponding simulated and measured transmission for both TM and TE polarizations of the incident THz wave. **(c)** Optical image of a polarizer realized using metallic gold instead of crystalline GeTe having the same dimension as P3 (P/w= 6/3 µm) and the associated transmission for TM and TE polarizations of the THz radiation.

Besides the very good agreement between the simulated and measured results for all the investigated designs, the GeTe-based fabricated devices show a broadband, relatively flat response over the entire investigated frequency domain. When comparing the results for the P1 and P3 polarizers we may notice that a decrease of the fill factor (defined as the ratio of the width of a single GeTe line to the period of the wire grid) from 0.3 (P1) to 0.2 (P3) for a fixed period (P= 10 µm) is decreasing the attenuation of the TE mode from -13.04 dB to -10.9 dB while the TM transmission is slightly improving from -0.26 dB for the P1 device to -0.22 dB for the P3 polarizer. The best performances were obtained for the P2 device (having a fill factor of 0.5 and a smaller period value, of P= 6 μm), showing an experimentally-recorded contrast of 16.5 dB between TE and TM polarizations. This behavior is consistent with previous studies on THz polarizers showing that fill factors over 0.5 should be preferred for achieving high extinction ratios between the TE and TM modes in the THz domain[50]. The polarization performances of the GeTe-based devices induced by laser irradiation are similar with those of identical devices in which the GeTe was prepared to the crystalline phase by direct thermal heating (supplementary material **Figure S6**).

By comparing the performance of the GeTe-based P2 polarizer to the device having the same design but fabricated with gold instead of crystalline GeTe, we clearly notice the better performance of the latter (P4 on **Figure 4c**) which is expected, taking into account the lower conductivity of the crystalline GeTe compared to that of Au. Nevertheless, a gold-based structure is a fixed and passive structure whereas the GeTe structure can be tuned between its amorphous and crystalline states by optical stimuli while showing very good polarization performances of the THz waves. Thus, we have demonstrated the feasibility to develop all dielectric (metal free) THz polarizers using solely chalcogenide phase change materials which presents a broadband response, a high extinction ratio when in metal-like phase (up to 16.5 dB) and almost transparent in the amorphous phase, using an optically- induced non-volatile transition.

To our knowledge, this is the first experimental demonstration of such broadband devices based on GeTe PCM in the THz domain. As the overall performances of such THz devices can be further improved through design adjustments, the presented approach is highly functional, technological disruptive and extremely stimulating for fast THz waves control and manipulation using reconfigurable or switchable THz devices such as frequency-selective surfaces, polarizers, lenses and waveplates, spatial modulators, power limiters etc. [45]

## 5. Conclusion

In conclusion, we have demonstrated hybrid and all-dielectric THz devices based on GeTe phase change material integration, which can be optically switched between its two different amorphous and crystalline phases, in a reversible way and with zero-power holding capabilities by using nanosecond UV laser pulses. We experimentally realized optically-induced non-volatile resonance switching by hybrid integration of GeTe with metallic elements within the unit cell of a metamaterial device. The good contrast of GeTe THz properties between the amorphous and the crystalline state, allowed us to design and prove for the first time the possibility to realize an all-dielectric reconfigurable THz devices based solely on GeTe which can achieve a strong modification of the polarization of incident THz waves (extinction ratios of 16.5 dB). The approach offers a great reconfiguration flexibility for multifunctional manipulation of THz waves, and can be easily adopted for implementing optical reconfiguration techniques capable to activate PCM structures within a particular THz device on large areas or on specific regions (through raster scanning or optical excitation of precise elements within the device) in order to program precise, on-demand functionalities of THz waves (arbitrary spatial profile of intensity, phase, and/or polarization distribution). This highly versatile approach that we are proposing, based on non-volatile optically controlled multi-operational THz devices integrating PCMs, will, most certainly, generate stimulating, disruptive developments in the THz domain like field- programmable metasurfaces, all-dielectric coding metamaterials with multifunctional capabilities for THz field manipulation.

## 6. Experimental section

The THz hybrid metamaterial was fabricated in a cleanroom environment by standard photolithography and deposition procedures using a two-mask levels process. Firstly, a 500-nm thick GeTe layer have been obtained on a 500-μm thick c-cut sapphire substrate using DC magnetron sputtering from a GeTe (50-50 stoichiometry) target in an argon atmosphere. The GeTe elements were patterned using a lithographically-defined photo-resist mask and a wet etching process. The subsequently deposited metallic patterns (20/500-nm thick Ti/Au bilayers) defining the SRR structure were obtained by electron-beam evaporation of the respective metallic elements and an optical lithography step using the lift-off method. The GeTe-based THz polarizer was obtained by patterning a GeTe film using a one-mask lithographically step defining a photo-resist mask and a subsequent dry etching process while the metallic gold-based polarizer was fabricated in a similar matter from a Ti/Au (20-500 nm) film obtained by electron-beam evaporation.

For the optical reversible switching of the GeTe film between the two amorphous and crystalline states we employed the direct irradiation of the obtained GeTe films using pulses from an excimer KrF laser with a wavelength of 248 nm and a pulse duration of 35 ns. Laser pulses with energy densities ~90 mJ/cm$^2$ where employed for transforming the GeTe from the amorphous to the crystalline state while the reverse transition requires energy densities in the range 185-190 mJ/cm$^2$. The conductivities of the GeTe films were measured using a four probes set-up.

The full-wave simulations of the THz devices were performed using the commercially available Ansys HFSS software package. For the GeTe materials parameters, we used the dc conductivities layer extracted in the two states by the four-probe method, while for the permittivity of the amorphous GeTe we used a value of 20. Similar permittivity values were reported previously for different PCM compositions [35].

The THz characterization of the bare GeTe films and of the different fabricated THz devices was realized using a homemade THz time domain spectroscopy (THz-TDS) system.

**Supporting Information**

Supporting Information is available from the authors.